# Unraveling the Effect of Electron-Electron Interaction on Electronic Transport in High-Mobility Stannate Films

Jin Yue[a], Laxman R. Thoutam, Abhinav Prakash, Tianqi Wang, and Bharat Jalan[a]

Department of Chemical Engineering and Materials Science

University of Minnesota,

Minneapolis, Minnesota 55455, USA

[a] Corresponding authors: yuexx129@umn.edu, bjalan@umn.edu

**Abstract**

Contrary to the common belief that electron-electron interaction (EEI) should be negligible in *s*-orbital-based conductors, we demonstrated that the EEI effect could play a significant role on electronic transport leading to the misinterpretation of the Hall data. We show that the EEI effect is primarily responsible for an increase in the Hall coefficient in the La-doped $SrSnO_3$ films below 50 K accompanied by an increase in the sheet resistance. The quantitative analysis of the magnetoresistance (MR) data yielded a large phase coherence length of electrons exceeding 450 nm at 1.8 K and revealed the electron-electron interaction being accountable for breaking of electron phase coherency in La-doped $SrSnO_3$ films. These results while providing critical insights into the fundamental transport behavior in doped stannates also indicate the potential applications of stannates in quantum coherent electronic devices owing to their large phase coherence length.

Alkaline-earth stannates with perovskite structures (BaSnO$_3$, SrSnO$_3$) have attracted significant research interest since the past decade due to their wide bandgap, high optical transparency and high room temperature mobility ($\mu_{RT}$).[1] These characteristics make them promising candidates for transparent electronics and power electronics applications.[2-7] Significant research effort has been devoted to BaSnO$_3$ (BSO) since the recent discovery of record-high $\mu_{RT}$ in perovskite oxides, which can reach up to 320 cm$^2$V$^{-1}$s$^{-1}$ in bulk[8] and 20-183 cm$^2$V$^{-1}$s$^{-1}$ in thin films when doped with La.[8-15] The close cousin of BSO, SrSnO$_3$ (SSO), has larger bandgap and smaller lattice parameter, yielding it more amenable to the heterostructure and strain engineering using commercially available substrates.[16] Critical progress has been made in synthesizing high-quality thin film of BSO and SSO in addition to characterizing their fundamental structure-properties relationships.[8-18] Practical devices operating at room temperature (MOSFET[2-5,19], MESFET[6], EDLT[20,21]) using both BSO and SSO have also been demonstrated.

Despite these remarkable progresses, there are still many open fundamental questions regarding the electronic transport in stannates. For example, the slight upturn in resistivity at low temperatures in degenerately doped stannates has received very little attention. These upturns are commonly attributed to the weak localization (WL) effect.[9,13] At the same time, a small but measurable drop in the nominal carrier density is also observed but often overlooked, or sometimes, even attributed to the carrier freeze-out. Understanding these effects is critically important for the utilization of these materials in practical applications such as high-mobility transistors or in the study of low-dimensional physics using high-mobility heterostructures.

About 40 years ago, Atshuler et. al. had predicted that the weak electron-electron interaction (EEI) effect can lead to an upturn in resistivity accompanied by an increase in the Hall coefficient at low temperatures.[22] Later, the same was demonstrated experimentally in Si, graphene and GaAs among others.[23-26] However, there are seldom any reports of the EEI effects in the *s*-orbital-based conductors. This is likely because *s*-orbitals are generally more dispersive than *p*, *d* and *f*-orbitals, and therefore, electron-electron interactions may be better screened, and the non-interacting-electron approximation stays valid.

In this letter, we show that the EEI effects can play a critical role on electronic transport at low temperatures in materials with *s*-orbital derived conduction band. Using La-doped SSO films, whose conduction band is mainly derived from Sn 5*s*-orbitals, we show that an increase in the Hall coefficient (or the decrease in the nominal carrier density) at low temperatures is due to the EEI effects and that it should be carefully accounted for in the interpretation of the Hall measurements. We also report on the fundamental materials parameters including record-high electron phase coherence length in La-doped SSO films.

30 nm La-doped SSO films were grown on 2 nm undoped buffer SSO on $GdScO_3$ (110) substrates using the hybrid molecular beam epitaxy (MBE) approach. Details of the growth and stoichiometry optimizations are described elsewhere.[16,18] For La-doping, La cell temperature was kept at 1150 °C. A high-resolution X-ray $2\theta–\omega$ couple scan of the structure is shown in figure 1a, confirming epitaxial and phase-pure SSO $(002)_{pc}$ film on GSO (110). Finite-size Kiessig fringes are observed, indicating smooth film morphology on a short lateral length scale. The out-of-plane lattice parameter calculated from SSO $(002)_{pc}$ peak was 4.115 Å ± 0.002 Å, which is close to the calculated value of the out-of-plane lattice parameter (4.119 Å) of a coherently strained tetragonal phase of SSO with *c*-axis in-plane on GSO (110) substrates. This result is consistent with the prior report on the epitaxial stabilization of tetragonal SSO on GSO (110).[16] Atomic force microscopy (AFM) image in figure 1b shows an atomically smooth surface morphology.

Temperature-dependent sheet resistance ($R_s$) is shown in figure 2a for 30 nm La-doped SSO/2 nm SSO/ GSO (110) revealing a linear upturn in $R_s$ vs. ln $T$ plot below 50 K (shown as a black dashed line) due to the quantum corrections. These quantum corrections to the classical Drude conductivity at low temperatures have two principal mechanisms: the 2D WL effect and the EEI effect. The former originates from the quantum self-interference of the electron partial waves scattered by impurities, while the latter comes from the interference of two different electrons with close energies (also referred to as Aronov-Altshuler effect).[27] While both these effects can lead to a similar ln $T$ correction to the longitudinal resistance ($R_{xx}$), only EEI produces ln $T$ correction to the Hall coefficient $R_H$ (= $-\rho_{xy}/B$). Furthermore, for EEI effect, the normalized correction to the Hall coefficient, $\Delta R_H / R_H$ is predicted to be twice as large as the normalized correction to the sheet resistance, $\Delta R_S / R_S$.[22] Mathematically, this can be expressed as $\frac{\Delta R_H}{R_H} = \gamma \frac{\Delta R_S}{R_S}$, where $\gamma = 2$ corresponds to the EEI effect; $\gamma = 0$ corresponds to the WL effect, and $0 < \gamma < 2$ indicates the coexistence of the EEI and WL effects.

To examine the governing mechanism(s) for the ln $T$ dependence in $R_s$, we performed Hall measurements at various temperatures. Results from the same sample, 30 nm La-doped SSO/2 nm SSO/ GSO (110) are shown in figure 2b indicating an obvious downturn in the nominal 3D carrier density ($-1/eR_H$) in the same temperature range (< 50 K) where an upturn in $R_s$ was observed in figure 2a. Such downturn can be due to the carrier freeze-out effect. An Arrhenius behavior between $-1/eR_H$ and $1/T$ can therefore be expected, which is not the case in our data as shown in the inset of figure 2b. Instead, we found that the decrease in the nominal carrier density, or the increase in $R_H$, is well described by a linear ln$T$ behavior as shown in figure 2c. This result suggests that the EEI effect is at play. To further examine whether or not the WL effect is present, we show in figure 2d, a plot of $\frac{\Delta R_H}{R_H}$ vs. $\frac{\Delta R_S}{R_S}$, along with a linear fit to the experimental data yielding a value of γ = 1.14. This result suggests the presence of WL in addition to the EEI effect.

For WL effect, the time reversal symmetry can be destroyed by applying weak magnetic field which breaks the phase coherence between two partial waves resulting in negative magnetoresistance, whereas for EEI effects the phase coherence is mainly limited by the temperature and less sensitive to weak magnetic field.[28,29] The simplified Hikami-Larkin-Nagaoka (HLN) formula[30] describes the low-field MR ($\frac{\Delta R_{WL}}{R}$) due to the 2D WL in the perpendicular magnetic field. According to the HLN formula,

$$\frac{\Delta R_{WL}}{R^2} = \frac{e^2}{2\pi^2\hbar}\left[\ln\left(\frac{\hbar}{4el_\varphi^2 B}\right) - \psi\left(\frac{\hbar}{4el_\varphi^2 B} + \frac{1}{2}\right)\right],$$

where $\psi$ is the digamma function, $B$ is the magnetic field, $\hbar$ is the reduced Plank constant, and $l_\varphi$ is the phase coherence length. It is noted that this formula is applicable in the low field, and in the quantum diffusive transport regime, (i.e. $l_\varphi$ must be greater than the elastic mean free path distance of electrons, $l_{el}$) and that the spin-orbit interactions are not accounted for. Significantly, there is only one fitting parameter, $l_\varphi$, which can be determined from the fitting of the experimental MR data. Furthermore, the temperature dependence of $l_\varphi$ allows determining the dephasing mechanism (s) and their dimensionality. For instance, it is mathematically established that $l_\varphi$ can scale with temperature as $l_\varphi \propto T^{-p/2}$ where the exponent $p$ depends on the dephasing mechanisms. If electron-electron interaction is the dominant dephasing mechanism (also called Nyquist dephasing mechanism), the value of $p$ is equal to 0.66, 1, 1.5 in 1D, 2D and 3D respectively.[28] These quasi-elastic electron-electron scatterings are equivalent to the interaction of an electron with the fluctuating electromagnetic field produced by all other surrounding electrons, i.e. dephasing by the equilibrium Nyquist noise.[31] If the dephasing is dominated by electron-phonon interaction, $p$ takes a value between 2 to 4.[31]

Now we turn to the discussion of experimental MR data and their analyses. Figure 3a shows low-field MR data (open symbols) as a function of temperature showing negative MR for $T < 50$ K. The negative MR is consistent with the presence of WL effect. The dotted lines in figure 3a show excellent fits to the experimental MR using the HLN formula as discussed above, which allows to not only determine $l_\varphi$ but also electron dephasing mechanism. The extracted value of $l_\varphi$ is plotted in figure 3b as a function of ln $T$ for T < 50 K along with a linear fit. This result indicates $l_\varphi > t_{film}$ (marked by a horizontal dashed line) for T < 50 K, which is again consistent with the 2D nature of the WL. Additionally, the linear fit to $\ln l_\varphi$ vs. ln $T$ plot yielded a value of $p = 1.36$ revealing 3D EEI effect being responsible for breaking the phase coherency of electrons, i.e. the dephasing mechanism. It is also noteworthy that the phase coherence length in this sample is 260 nm at 1.8 K, which increases to 460 nm at 1.8 K with increasing carrier density ($n_{3D} = 5.5 \times 10^{19}$ cm$^{-3}$) (see figure S1). Significantly, this value is twice as large as the largest value reported for the doped SrTiO$_3$ at similar temperatures[32,33] indicating potential applications of doped SSO films in quantum coherent devices. We also calculated the electron-electron interaction characteristic length $l_{ee}$, using $l_{ee} = \sqrt{\dfrac{D\hbar}{2\pi k_B T}}$, where $D$ is the diffusion constant of electrons[29] (see SI). The corresponding $l_{ee}$ as a function temperature is plotted in figure 3b for T < 50 K which confirms that the EEI interaction is effectively 3D is our system, which is again consistent with the above results based on the value of $p$.

In summary, we have revealed the "unexpected" electron-electron interaction effect in La-doped SSO films. Both WL and EEI effects were found to be responsible for the low temperature quantum correction in sheet resistance. A measurable increase in $R_H$ below 50 K was found due to the EEI effects and therefore, should not be attributed to a decrease in carrier density. The MR results were fitted to the 2D WL theory, and the phase coherent length $l_\varphi$ was extracted from the fitting. From the temperature dependence of the $l_\varphi$, the dephasing mechanism was determined to be 3D EEI effects. At 1.8 K, a large phase coherence length of 460 nm was obtained, suggesting potential applications of La-doped SSO in quantum coherent devices.

**Acknowledgements:**

The authors thank A. Kamenev for helpful discussion. This work was primarily supported by the Young Investigator Program of the Air Force Office of Scientific Research (AFOSR) through Grant FA9550-16-1-0205. Part of this work was supported through UMN MRSEC program under Award Number DMR-1420013 and through DMR-1741801 and DMR-1607318. Parts of this work were carried out at the Minnesota Nano Center, which is supported by the National Science Foundation through the National Nano Coordinated Infrastructure (NNCI) under Award Number ECCS-1542202. Structural characterizations were carried out at the University of Minnesota Characterization Facility, which receives partial support from NSF through the MRSEC program.

Figures: (Color online)

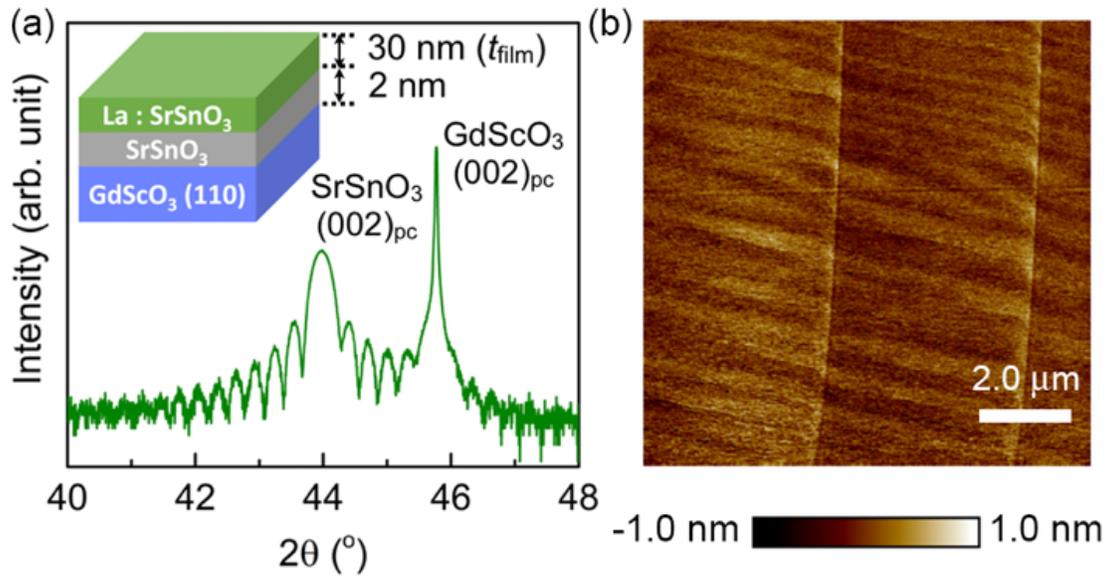

**Figure 1:** (a) High-resolution X-ray 2θ-ω couple scan of the 30 nm La-doped SSO/2 nm undoped SSO film grown on GSO (110) substrate. The schematic of the structure is shown in the inset. (b) Atomic force microscopy (AFM) image of the sample, showing a smooth surface morphology.

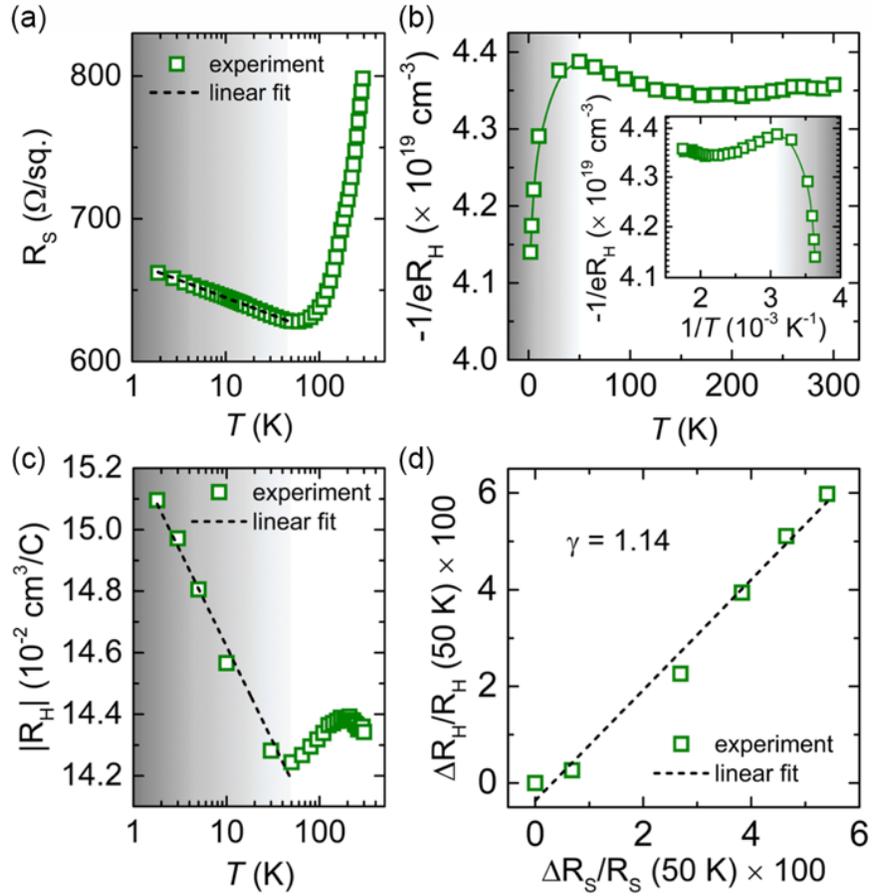

**Figure 2**: (a) $R_s$ vs. ln $T$ plot of a 30 nm La-doped SSO/2 nm undoped SSO film grown on GSO, showing a ln $T$ correction below 50 K, (b) $-1/eR_H$ as a function of temperature where $R_H$ is the Hall slope. Inset shows an Arrhenius plot (ln $(-1/eR_H)$ vs. $1/T$) indicating no linear relationship for T < 50 K, (c) $|R_H|$ vs. ln $T$ plot with a linear fit showing a linear ln $T$ behavior below 50 K, (d) normalized correction to $R_H$ as a function of normalized correction to $R_s$, showing a linear relationship with a slope $\gamma$ = 1.14. Dotted lines are linear fits to the data. The temperature range (T < 50 K) where quantum correction is operative is marked by a color gradient for clarity.

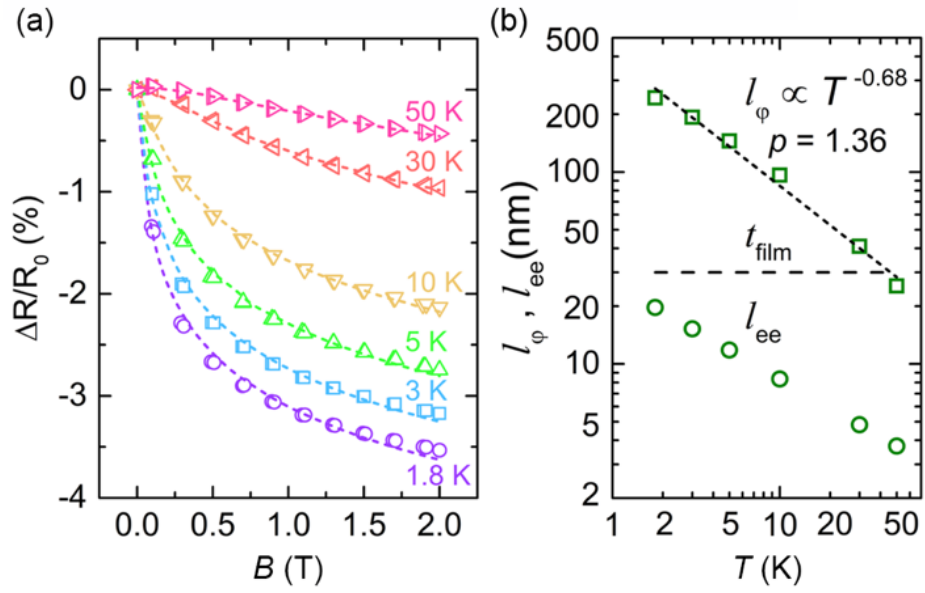

**Figure 3**: (a) Normalized MR as a function of temperature and magnetic field with fitting (dotted lines) using the 2D WL theory, (b) log-log plot for the extracted phase coherence length $l_\varphi$ (squares), active film thickness $t_{film}$ (horizontal dash lines), and the characteristic length for EEI (circles) $l_{ee}$ as a function of temperature. Note the dashed line is a fit to the $\ln(l_\varphi)$ vs. $\ln T$ yielding $p = 1.36$.